\documentclass[%
 aps, prl, letterpaper, floatfix,
 preprintnumbers,
 reprint,
 amsmath,amssymb,
 showpacs,
]{revtex4-1}

\usepackage{graphicx}
\usepackage{bm}
\usepackage[breaklinks=true]{hyperref}
\hypersetup{bookmarksnumbered, pdfpagemode=UseOutlines, 
pdfauthor={I.\ J.\ Vera-Marun}, 
pdftitle={Nonlinear interaction of spin and charge currents in graphene}, 
pdfdisplaydoctitle, 
colorlinks=true, citecolor=blue, filecolor=blue, linkcolor=blue, urlcolor=blue}

\begin{document}

\title{Nonlinear interaction of spin and charge currents in graphene}

\author{I.\ J.\ \surname{Vera-Marun}}
	\email[Electronic address: ]{i.j.vera.marun@rug.nl} 
\author{V.\ Ranjan}
\author{B.\ J.\ \surname{van Wees}}
\affiliation{Physics of Nanodevices, Zernike Institute for Advanced Materials, University of Groningen, The Netherlands}

\date{\today}

\begin{abstract}
We describe a nonlinear interaction between charge currents and spin currents which arises from the energy dependence of the conductivity. This allows nonmagnetic contacts to be used for measuring and controlling spin signals. We choose graphene as a model system to study these effects and predict its magnitudes in nonlocal spin valve devices. The ambipolar behavior of graphene is used to demonstrate amplification of spin accumulation in \emph{p-n} junctions by applying a charge current through nonmagnetic contacts.
\end{abstract}

\pacs{72.25.Hg, 72.80.Vp, 75.76.+j, 85.75.-d}
\preprint{Resubmitted on December 9, 2011}

\maketitle

Spin-polarized transport (spintronics) \cite{uti_spintronics:_2004} in graphene, a one-atom-thick layer of carbon \cite{castro-neto_electronic_2009}, is of both fundamental and technological interest due to its long spin relaxation length \cite{tombros_electronic_2007} and large spin signals \cite{han_tunneling_2010}. In this work we focus on understanding graphene spintronics as it is experimentally addressed by an all-electrical scheme involving the use of a nonlocal device geometry \cite{tombros_electronic_2007}. The conductivity in graphene has been considered only at the Fermi level, which leads to equal values for both spin channels. We point out that very recent work \cite{abanin_giant_2011-1} uses the energy dependent conductivity of graphene, in conjunction with Zeeman splitting by applied magnetic fields, to explain a giant spin-Hall effect in the linear regime.

In this contribution we highlight \emph{nonlinear} effects in the absence of external magnetic field that gives rise to interactions between electron spin and charge in graphene. We argue that the nonlinear interaction between spin and charge can be consistently described by using the equations for spin diffusion in metals \cite{van_son_boundary_1987,valet_theory_1993}, while considering the energy dependent conductivity $\sigma(\epsilon)$ of graphene and the large spin accumulation achievable by spin injection. This gives rise to phenomena observable in the nonlocal geometry, which include detection of spin accumulation in graphene via nonmagnetic contacts, its manipulation using charge currents and amplification in bipolar devices.

Previous experimental work has shown the manipulation of spin accumulation in graphene by applying high electric fields \cite{jozsa_electronic_2008,*jozsa_controlling_2009}. Such a manipulation has been later interpreted by considering the effect of low-resistance contacts \cite{yu_transfer_2010} within the drift-diffusion formalism for spin accumulation derived for semiconductors \cite{yu_electric-field_2002,*yu_spin_2002}. In the following, we consider highly resistive noninvasive contacts \cite{schmidt_fundamental_2000,rashba_theory_2000}, which can be treated as ideal (spin) current injectors and (spin) voltage detectors \cite{popinciuc_electronic_2009,han_tunneling_2010}. This allows us to only focus on the physics of spin transport \emph{within} graphene.

We start with the well established model for spin transport in metals \cite{van_son_boundary_1987,valet_theory_1993} with the electrochemical potential for each spin channel expressed as $\mu_\pm=\mu_{\text{avg}}\pm\Delta\mu$, where the index $\pm$ refers to electron spin $\pm1/2$ . Here, $\Delta\mu$ is the term related to the spin accumulation and $\mu_{\text{avg}}$ the average potential. This results in spin-diffusion equations
\begin{eqnarray}
\frac{\partial^2 \Delta\mu}{\partial x^2}&=&\frac{\Delta\mu}{\lambda^2}\;, \label{eq:spindiff}\\
J_\pm(x)&=&\sigma_\pm \left[ E(x)\pm\frac{1}{e}\frac{\partial \Delta\mu}{\partial x} \right] \;, \label{eq:spincurr}
\end{eqnarray}
with $e$ the electron charge, $\lambda$ the spin relaxation length, $\sigma_\pm$ and $J_\pm$ the conductivities and the currents for each spin channel, and $E(x)=(1/e)(\partial \mu_{\text{avg}}/\partial x)$.
As shown in Eq.~(\ref{eq:spincurr}), the gradient in $\Delta\mu$ drives the current for each spin channel in opposite directions, whereas the electric field $E$ drives them in the same direction.
The spin-dependent conductivities are described by the polarization $\beta$ such that $\sigma_\pm=[2\rho(1\pm\beta)]^{-1}$. The general solutions for $\mu_\pm$, $J_\pm$, and $E$ in a homogeneous medium were presented in Appendix C of Ref.~\onlinecite{valet_theory_1993}. To find numerical solutions for a certain device configuration we divide the graphene into discrete regions and use the solutions for each region, while keeping continuities of $\mu_\pm$ and $J_\pm$.

\begin{figure}[tbp] 
\includegraphics*[angle=0, width=0.5\textwidth]{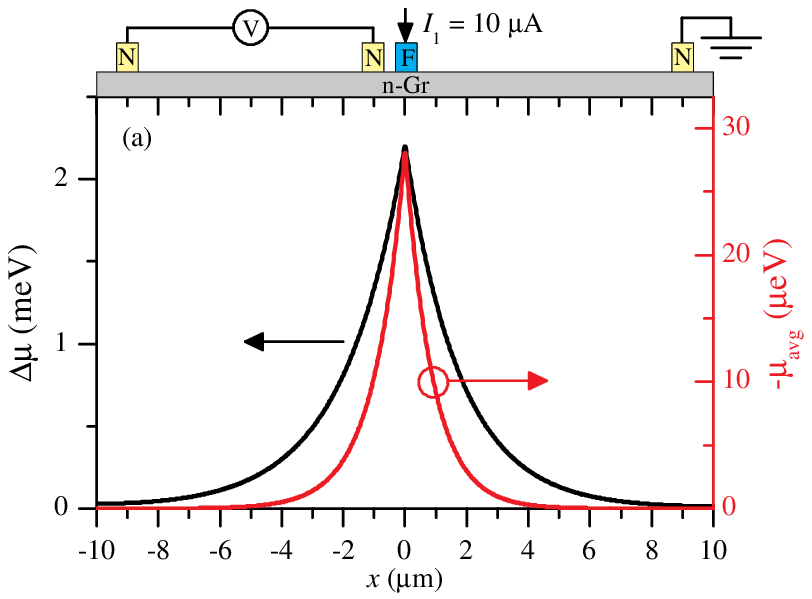}
\includegraphics*[angle=0, width=0.5\textwidth, trim = 0mm 0mm 0mm 6mm, clip]{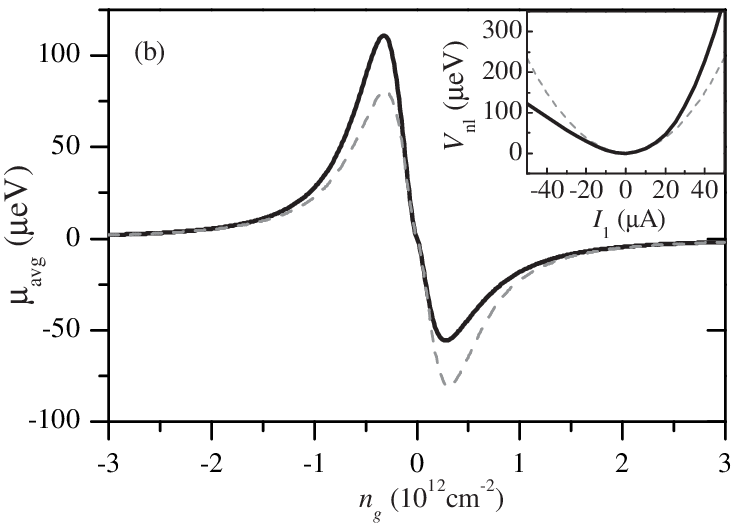}
\caption{\label{fig:one}
Generation of a nonlocal charge voltage $V_{\text{nl}} = -\mu_{\text{avg}}/e$ by spin injection in graphene. (a) Profiles of $\mu_{\text{avg}}$ and $\Delta\mu$ created by a spin current. The linear ohmic drop due to $I_1$ on the right side of the circuit has been subtracted from $\mu_{\text{avg}}$ for the sake of clarity. (b) Charge carrier density dependence of the nonlocal potential $1~\mu m$ away from the injector. The inset shows $V_{\text{nl}}$ measured by a nonmagnetic contact versus $I_1$. Dashed curves correspond to injection of pure spin current $PI_1$ without a charge current, whereas solid curves are for considering the effect of a charge current $I_1$ in the right side of the circuit.
}
\end{figure}

In metals it is possible to achieve a spin accumulation $\Delta\mu \approx10$~$\mu$eV \cite{jedema_electrical_2002} whereas in graphene it can be considerably larger, as explained below. Electrical spin injection from a tunnel contact with polarization $P$ using a charge current $I_1$ results in injection of a spin current $P I_1$. Ignoring any interaction between spin and charge, $\Delta\mu$ decays away from the injector as given by \cite{tombros_electronic_2007}
\begin{eqnarray}
\Delta\mu=\frac{P \rho \lambda e I_1}{2 w}\exp{\left(-\frac{|x|}{\lambda}\right)} = e I_1 \frac{R_s}{P}\; , \label{eq:std}
\end{eqnarray}
with $\rho=1/\sigma$ the graphene square resistance, $w$ its width, and $R_s$ the nonlocal spin resistance as measured by a second magnetic contact with same $P$. For typical values of $\rho\approx1$~k$\Omega$, $w=1~\mu$m, $\lambda=2~\mu$m, $I_1=10~\mu$A and $P=20$~\% the resulting $\Delta\mu$ profile reaches $\approx1$~meV, as depicted in Fig.~\hyperref[fig:one]{\ref*{fig:one}(a)}. At such large splitting in the electrochemical potential the energy dependence of the graphene conductivity $\sigma(\epsilon)$ starts to be noticeable. Therefore we introduce a splitting dependent $\beta$ given by
\begin{eqnarray}
\beta\approx-\left.\frac{1}{\sigma}\frac{\partial \sigma}{\partial \epsilon}\right|_{\mu_{\text{avg}}} \hspace{-2.5ex} \Delta\mu = -\alpha \: \Delta\mu\;, \label{eq:alpha}
\end{eqnarray}
with a conductivity spin polarization proportional to $\Delta\mu$ and a coefficient $\alpha$. The effects of temperature and disorder, described later, can be taken into account by considering the experimental electrical conductivity $\sigma(\epsilon,T)$.

Now, we discuss the nature of the coefficient $\alpha$. The conductivity of graphene away from the neutrality (Dirac) point can be described by $\sigma=\nu n e$, with $\nu$ the carrier mobility and $n$ the carrier density. Due to the linear density of states in graphene \cite{castro-neto_electronic_2009} the carrier density depends on energy as $n(\epsilon) = \epsilon^2 / \pi \hslash^2 v_F^2$, where $\hslash$ is the reduced Planck constant and $v_F \approx 10^6$~m/s is the Fermi velocity. For a constant mobility, the latter leads to $\alpha=2/\epsilon \propto 1/\sqrt{n}$, so $\alpha$ diverges when $n$ tends to zero. Therefore the maximum value of $\alpha$ is given by the mechanisms limiting how close we can get to $n=0$. The charge carrier density in graphene field-effect transistors is electrostatically tunable by applying a voltage $V_g $ to a gate such that $n_g=(C_g/e)(V_g - V_d)$ with $C_g$ the gate capacitance per unit area and $V_d$ the voltage at which the neutrality point is located. A background carrier density $n_i$ is induced by the presence of electron-hole puddles ($n_{pd}$) due to disorder \cite{martin_observation_2008} and thermal generation of carriers ($n_{th}$) \cite{dorgan_mobility_2010}. Up to date, all spintronic devices have been fabricated with graphene supported on a substrate, mostly SiO$_2$. At room temperature, for a typical value of $n_i = (n_{pd}^2 + 4n_{th}^2)^{1/2} = 4\times10^{11}$~cm$^{-2}$, we obtain a maximum value of $\alpha\approx (60$~meV)$^{-1}$ which together with $\Delta\mu\approx1$~meV yields $\beta\approx1$\%.

We use the modeling above to study spin transport in graphene field-effect transistors. We can describe most experimental $\sigma$ vs $V_g$ (Dirac) curves for graphene on SiO$_2$ by taking a constant value of $\nu=0.4$~m$^2$/Vs and a carrier density $n=(n_g^2+n_i^2)^{1/2}$ \cite{dorgan_mobility_2010}. For simplicity, we keep the polarization of magnetic contacts fixed at $P=20$~\% and a carrier density independent spin relaxation length of $\lambda=2~\mu$m. Unless stated otherwise, we use $n_g=2\times10^{12}$~cm$^{-2}$ [$\alpha\approx (83$~meV)$^{-1}$], well into the metallic regime. First, we consider in Fig.~\hyperref[fig:one]{\ref*{fig:one}(a)} a nonlocal measurement where a spin current $PI_1$ is injected into graphene. The presence of a spin current in graphene with an inhomogeneous conductivity polarization $\beta$ creates a nonlocal charge voltage $V_{\text{nl}} = -\mu_{\text{avg}}/e$. Interestingly, such a potential is detectable with both magnetic and nonmagnetic contacts. For the simple case of pure spin current injection (ignoring the effect of the charge current $I_1$ on the right side of the circuit) we obtain
\begin{eqnarray}
V_{\text{nl}} = \frac{\alpha}{2 e}(\Delta\mu)^2 = \frac{\alpha e}{2}\left(\frac{R_s}{P}\right)^2I_1^2 \; , \label{eq:Vnl}
\end{eqnarray}
which indicates that this is a second order effect in $\Delta\mu$. The latter also means that $V_{\text{nl}}$ must decay with a characteristic length of $\lambda/2$. The nonlinear nature of $V_{\text{nl}}$ is explicit in the inset of Fig.~\hyperref[fig:one]{\ref*{fig:one}(b)}. This effect opens the unique possibility of measuring spin signals in graphene without using magnetic detectors.

Next, we consider the effect of changing the graphene carrier density $n_g$ on $V_{\text{nl}}$. We calculate $\alpha$ from Eq.~(\ref{eq:alpha}) using the simulated electrical conductivity $\sigma(\epsilon,T)$, which includes the effect of $n_i$, such that $\partial \sigma / \partial \epsilon = (\partial \sigma / \partial n_g) (\partial n_g / \partial \epsilon)$. The coefficient $\alpha$ has opposite polarity for hole and electron regimes. Besides, a finite $n_i$ introduces the presence of both electrons and holes near the Dirac point, which removes the divergence of $\alpha$ and makes it zero. This behavior is in analogy to the case of the thermoelectric Seebeck coefficient $S$ which has the same dependence on $\sigma(\epsilon)$ as $\alpha$ and follows an approximate Mott formula based on the experimental Dirac curve \cite{laefwander_impurity_2007,*zuev_thermoelectric_2009}. From Eqs.~(\ref{eq:std}), (\ref{eq:alpha}) and (\ref{eq:Vnl}) the dependence of $V_{\text{nl}}$ on energy is given by $\sigma^{-3} \partial \sigma / \partial \epsilon$.

\begin{figure}[tbp]
\includegraphics*[angle=0, width=0.5\textwidth]{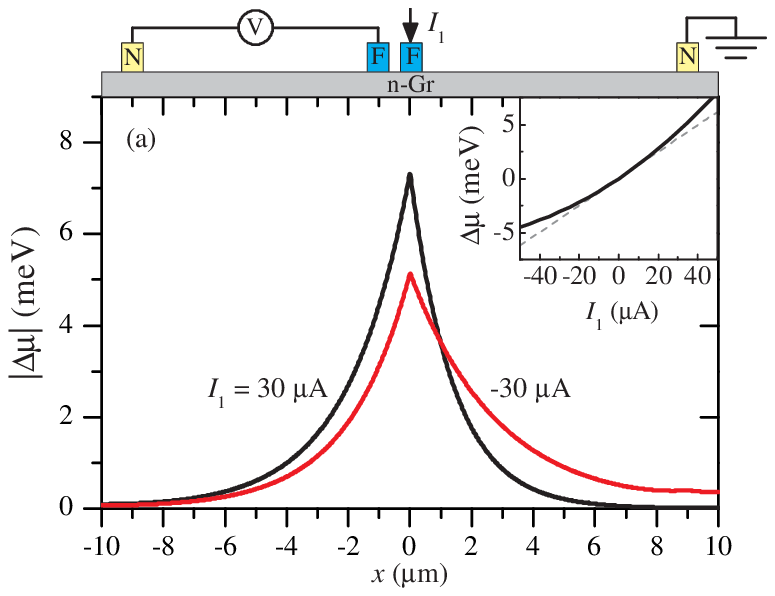}
\includegraphics*[angle=0, width=0.5\textwidth]{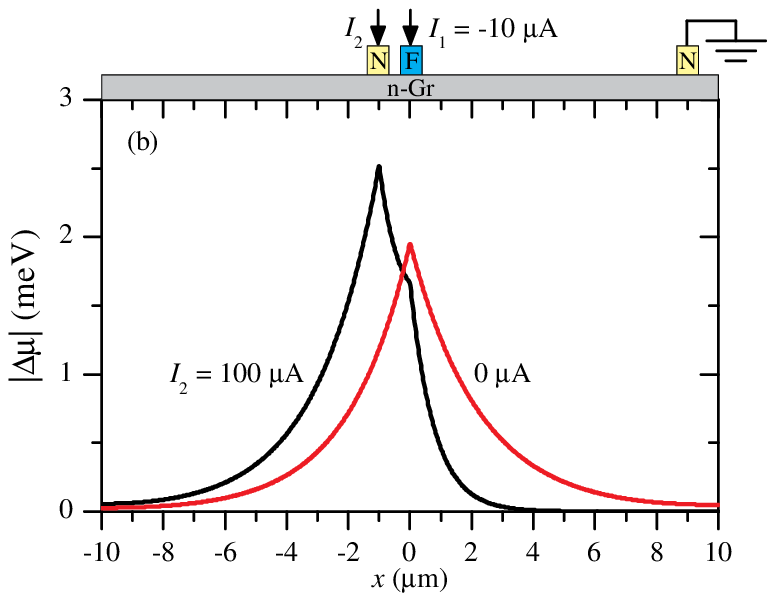}
\caption{\label{fig:two}
Effect of charge current on spin accumulation $\Delta\mu$. (a) Profile of $\Delta\mu$ due to its interaction with $I_1$ on the right side of the circuit. Inset: nonlinearity of the nonlocal $\Delta\mu$, $1~\mu$m away from the injector. (b) Redistribution of $\Delta\mu$ caused by a current $I_2$ via a nonmagnetic contact.
}
\end{figure}

The nonlocal charge voltage goes as $V_{\text{nl}} \propto I_1^2$ for injection of pure spin currents. If we also consider the charge current $I_1$ on the right side of the circuit [Fig.~\hyperref[fig:one]{\ref*{fig:one}(a)}], an asymmetry in the $V_{\text{nl}}$ vs $I_1$ curve is visible [inset in Fig.~\hyperref[fig:one]{\ref*{fig:one}(b)}]. This is a higher ($\geq 3^\text{rd}$) order effect on $V_{\text{nl}}$ given by the interaction of $I_1$ with the spin accumulation on the right side of the circuit, which creates a higher ($\geq 2^\text{nd}$) order effect on the nonlocal spin accumulation $\Delta\mu$. We make explicit such an effect on $\Delta\mu$ in Fig.~\hyperref[fig:two]{\ref*{fig:two}(a)}. The resulting nonlocal spin resistance detected in a spin valve with a second magnetic contact, $R_s = P \Delta\mu / e I_1$, now varies with the injection current, as observed in the nonlinear behavior shown in the inset of Fig.~\hyperref[fig:two]{\ref*{fig:two}(a)}. 

An interesting result is observed if we consider a second charge current $I_2$ via a nonmagnetic contact, as depicted in Fig.~\hyperref[fig:two]{\ref*{fig:two}(b)}. A spin accumulation in graphene creates a conductivity spin polarization $\beta$, which in the presence of a charge current $I_2$ gives rise to a spin current $\beta I_2$. The nonmagnetic contact hence seems to inject a spin current similarly to the case of a magnetic contact. Depending on the polarities of $I_2$ and $\alpha$, spin accumulation or depletion is observed. The nonmagnetic contact offers an extra handle by which we can amplify the spin accumulation. As shown in Fig.~\hyperref[fig:two]{\ref*{fig:two}(b)}, $\Delta\mu$ under the nonmagnetic contact can be even larger than that under the magnetic contact.

In graphene field-effect devices we can individually address specific regions via local electrostatic gates. We use this capability to study ambipolar spin transport in graphene. We choose a highly symmetric case where the physics can be easily understood and derive a simple analytical description which accurately describe the simulations. The latter is possible because in graphene we can ignore the effects of the charge depletion region present in nondegenerate \emph{p-n} junctions \cite{uti_spintronics:_2004, pershin_focusing_2003}.

\begin{figure}[tbp]
\includegraphics*[angle=0, width=0.5\textwidth]{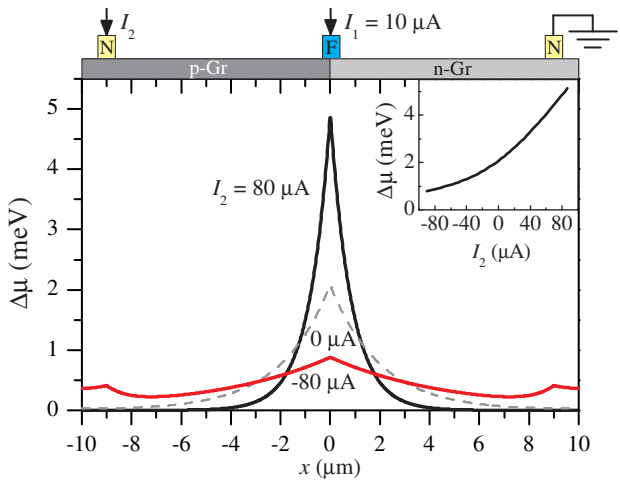}
\caption{\label{fig:three}
Change in spin accumulation $\Delta\mu$ profile in a graphene \emph{p-n} junction due to a charge current $I_2$ (assuming pure spin current injection $P I_1$). Inset: amplification of $\Delta\mu$ at the junction (under the magnetic contact) due to $I_2$.
}
\end{figure}

We consider a graphene channel with the left half set in the hole regime and the right half set in the electron regime, as depicted in Fig.~\ref{fig:three}. Initially, with $I_2=0$, the magnetic contact located at the junction creates a spin accumulation $\Delta\mu_0$ given by Eq.~(\ref{eq:std}) (assuming pure spin current injection). We define $A=\lambda\rho e/2w$, so that under the magnetic contact we have an initial $\Delta\mu_0=A P I_1$. If we now apply a charge current $I_2$ via the nonmagnetic contacts, the sign change of the parameter $\alpha$ at the junction creates a source of spin current equal to $2 \beta I_2$. Such a discontinuity induces a spin accumulation $\Delta\mu_{\text{ind}}$. We remark that the graphene spin polarization $\beta$ is given by the \emph{total} spin accumulation at the junction $\Delta\mu_{\text{tot}} = \Delta\mu_{0} + \Delta\mu_{\text{ind}}$. Therefore at the magnetic contact
\begin{eqnarray}
\Delta\mu_{\text{tot}}&=&A P I_1 + A 2\alpha I_2 \Delta\mu_{\text{tot}} + \xi\;, \label{eq:utot}\\
\xi&=&A 2\alpha I_2 \int^{\infty}_{0} \left[ \frac{\partial \Delta\mu_{\text{tot}}(x)}{\partial x} \exp{\left(-\frac{x}{\lambda}\right)} \right] dx \;, \label{eq:xi}
\end{eqnarray}
with $\xi$ a compensation term in $\Delta\mu_{\text{ind}}$ corresponding to the 
presence of $I_2$ in the $p$ and $n$ regions with inhomogeneous spin polarization $\beta$. For small charge currents ($I_2 \ll 1/\alpha A$) we have $\Delta\mu_{\text{tot}}(x) \approx \Delta\mu_0(x)$ and the integral in Eq.~(\ref{eq:xi}) evaluates to $-\Delta\mu_{\text{tot}}/2$. Introducing this result into Eq.~(\ref{eq:utot}) leads to $\Delta\mu_{\text{tot}}$ at the junction
\begin{eqnarray}
\Delta\mu_{\text{tot}}&=&\frac{A P I_1}{1 - A \alpha I_2} \;, \label{eq:feedback}
\end{eqnarray}
equivalent to an amplifier circuit with positive feedback controlled by $\alpha I_2$. Eq.~(\ref{eq:feedback}) gives accurate results for low values of $I_2$. The divergence at $A\alpha I_2=1$ is a result of our approximation for $\xi$. In reality, the distribution of spin accumulation will (de)focus at the junction with changing $I_2$ \cite{pershin_focusing_2003} which yields different compensation $\xi$.

To account for large values of $I_2$ we generalize Eqs.~(\ref{eq:utot}) and (\ref{eq:xi}) for the case of spin accumulation $\Delta\mu_{\text{tot}}(x)$ at any location within the graphene. It follows that the general solution of $\Delta\mu_{\text{tot}}$ satisfies
\begin{eqnarray}
\frac{\partial^2 \Delta\mu}{\partial x^2} + \frac{A2\alpha I_2}{\lambda}\frac{\partial \Delta\mu}{\partial x}-\frac{\Delta\mu}{\lambda^2}=0 \;, \label{eq:diff-utot}
\end{eqnarray}
where the second term arises due to the spin-dependent conductivity. We describe electronic transport in energy space via $\alpha$. For a one-dimensional Drude model the mathematical formulation is similar to that of drift \cite{yu_electric-field_2002}.

Eq.~(\ref{eq:diff-utot}) has solutions of the form $\exp(\mp x/L_{\pm})$ with $L_\pm = \lambda /(\pm A\alpha I_2 + \sqrt{(A\alpha I_2)^2+1})$. Using these solutions together with Eqs.~(\ref{eq:utot}) and (\ref{eq:xi}) we find that, for the case of nonmagnetic contacts far away from the spin injector, the spin accumulation has the form 
$\Delta\mu_{\text{tot}}(x) = (API_1/\lambda) L_- \exp{(-|x|/L_+)}$. The analytical solution describes the (de)focusing of the $\Delta\mu$ profile with $I_2$ shown in Fig.~\ref{fig:three}. At $I_2>0$ the distribution of $\Delta\mu$ focuses at the junction. The opposite occurs for $I_2<0$. In the limit $I_2 \gg 0$ the peak in the spin accumulation has a value of $\Delta\mu = 2A^2P \alpha I_1 I_2$ and the distribution tends towards a Dirac delta function with constant area $API_1\lambda$. The small peaks at $x = \pm 9$~$\mu$m are due to amplification at the nonmagnetic contacts. The dependence of $\Delta\mu$ at the junction versus $I_2$ is shown in the inset of Fig.~\ref{fig:three}. Finally, note that using the solutions to Eq.~(\ref{eq:diff-utot}) we can also describe the nonlinear spin resistance caused by $I_1$ in Fig.~\hyperref[fig:two]{\ref*{fig:two}(a)}, as $\Delta\mu=(P \rho \lambda_{\text{eff}} e I_1)(2 w)^{-1}\exp(-|x|/\lambda)$ for $x\leq0$, where $\lambda_{\text{eff}} = 2(1/\lambda + 1/L_-)^{-1}$. For small $I_1$ the latter leads to the addition of a second order term to Eq.~(\ref{eq:std}) of the form $V_{\text{nl}} e/P$, with $V_{\text{nl}}$ defined in Eq.~(\ref{eq:Vnl}).

In conclusion, we have described the interaction between spin and charge transport in graphene by treating it as a ferromagnet with a conductivity spin polarization $\beta$ induced by the presence of a spin accumulation $\Delta\mu$. This leads to phenomena experimentally accessible via nonlocal measurements, including detection and manipulation of spin signals with nonmagnetic contacts, its dependence on carrier density and amplification effects in ambipolar devices. Since the nonlinear interaction arises solely due to the energy dependence of the conductivity, the ideas described in this work are also applicable to other materials used for spin transport, such as Si and GaAs. The generality of this interaction is analogous to the interaction between heat and charge described by thermoelectricity.

\begin{acknowledgments}
We thank N.~Tombros and T.~Maassen for critically reading the manuscript. IJVM thanks C.~J\'{o}zsa for useful discussions. This work was financed by the Zernike Institute for Advanced Materials.
\end{acknowledgments}


%

\end{document}